\documentclass[twocolumn, amsmath, twocolappendix]{aastex631}
\usepackage{amsmath}

\begin{document}

\title{Ultra-high-energy Cosmic Ray Sources can be Gamma-ray Dim  }

\author{Angelina Sherman}
\affiliation{Department of Physics, Wisconsin IceCube Particle Astrophysics Center, University of Wisconsin, Madison, WI, 53706}

\author{Ke Fang}
\affiliation{Department of Physics, Wisconsin IceCube Particle Astrophysics Center, University of Wisconsin, Madison, WI, 53706}

\author{Rafael \surname{Alves Batista}}
\affiliation{Instituto de F\'isica Te\'orica UAM/CSIC, Calle Nicol\'as Cabrera 13-15, Cantoblanco, 28049 Madrid, Spain}
\affiliation{Sorbonne Universit\'e, CNRS, UMR 7095, Institut d’Astrophysique de Paris, 98 bis bd Arago, 75014 Paris, France}

\author{Rogerio Menezes \surname{de Almeida}}
\affiliation{Instituto de F\'isica, Universidade Federal do Rio de Janeiro, Caixa Postal 68528, Rio de Janeiro, RJ 21941-972, Brazil}

\date{\today}

\begin{abstract}

Ultra-high-energy cosmic rays, accelerated hadrons that can exceed energies of $10^{20}$ eV, are the highest-energy particles ever observed. While the sources producing UHECRs are still unknown, the Pierre Auger Observatory has detected a large-scale dipole anisotropy in the arrival directions of cosmic rays above 8 EeV. In this work, we explore whether resolved gamma-ray sources can reproduce the Auger dipole.  We use various {\it Fermi} Large Area Telescope catalogs as sources of cosmic rays in CRPropa simulations. We find that in all cases, the simulated dipole has an amplitude significantly larger than that measured by Auger, even when considering large extragalactic magnetic field strengths and optimistic source weighting schemes. Our result implies that the resolved gamma-ray sources are insufficient to account for the population of sources producing the highest-energy cosmic rays, and there must exist a population of UHECR sources that lack gamma-ray emission or are unresolved by the current-generation gamma-ray telescopes.

\end{abstract}

\section{Introduction}

Ultra-high-energy cosmic rays (UHECRs), charged nuclei whose energies can exceed $10^{20}$ eV, are the highest-energy particles ever observed. The sources of these particles are still unidentified, and the mechanism by which these extremely energetic particles are accelerated remains a mystery \citep{Kotera_2011, Alves_Batista_2019}. 
The Pierre Auger Observatory (Auger) has detected a dipole anisotropy in cosmic ray arrival directions above 8 EeV. The measured dipole has an amplitude of 6.5\% and points in a direction $\sim 125^{\circ}$ away from the Galactic center, supporting a hypothesis of extragalactic origin for the highest-energy cosmic rays. \citep{Auger_dipole_2017, Auger_dipole_2018, Auger_dipole_2020, Auger_dipole_2023}. 

Previous cross-correlation studies have indicated some possible excesses of UHECRs around gamma-ray producing sources. For example, the Pierre Auger Observatory recently detected a hotspot in the region of Centaurus A, NGC 4945, and M83 \citep{Abreu_2022}. In addition, the TA hotspot for UHECRs $>$ 57 EeV \citep{TA_hotspot} in the northern sky lies in the region of a few nearby gamma-ray emitting sources, in particular Mkn~421, Mkn~180, and M82.
Cross-correlation studies between gamma-ray bright sources and UHECR events also found an excess of cosmic ray events around starburst galaxies \citep{starburst_correlation}. 

Motivated by these observations, our work attempts to address whether the population of gamma-ray sources could encapsulate the population of UHECR-producing sources. We use extragalactic sources from the {\it Fermi} Large Area Telescope (LAT) gamma-ray source catalogs as the origins of cosmic rays in simulations with CRPropa3 \citep{4FGL, 3FHL, 4LAC, crpropa, crpropa3.2}.
We conclude that the known gamma-ray sources are insufficient to reproduce a dipole as low as that observed by Auger, suggesting that there may exist a population of gamma-ray dim UHECR sources.

In Section \ref{gr_sources} we discuss properties of gamma-ray catalogs with {\it Fermi}-LAT 4FGL-DR4 as a benchmark case, as well as preprocessing measures applied to the catalog before simulation. In Section \ref{simulation} we describe simulations of UHECRs with CRPropa3, especially our approach to accounting for deflections in the extragalactic and Galactic magnetic fields. In Section \ref{results} we present our simulation results, and in Section \ref{summary} we discuss the implication of our results.

\section{Properties of the {\it Fermi}-LAT 4FGL-DR4 catalog \label{gr_sources}}

The Large Area Telescope (LAT) aboard the {\it Fermi} Satellite surveys the sky in the GeV energy range, and has produced several generations of gamma-ray source catalogs, including the LAT 14-year Source Catalog (4FGL-DR4; \citealp{4FGL_DR4, 4FGL}),  Third {\it Fermi}-LAT Catalog of High-Energy Sources (3FHL; \citealp{3FHL}), and the Fourth Catalog of Active Galactic Nuclei  (4LAC-DR3; \citealp{4LAC}). Below we use 4FGL-DR4, which is the latest catalog from {\it Fermi}-LAT in the 50~MeV-1~TeV energy range as the basis for our simulations. 
We present results using the other GeV gamma-ray catalogs, which are consistent with those using 4FGL-DR4, in Appendix~\ref{appendix:othergammaCat}.

\begin{figure}[t]
    \centering
    \includegraphics[width = 0.9\linewidth]{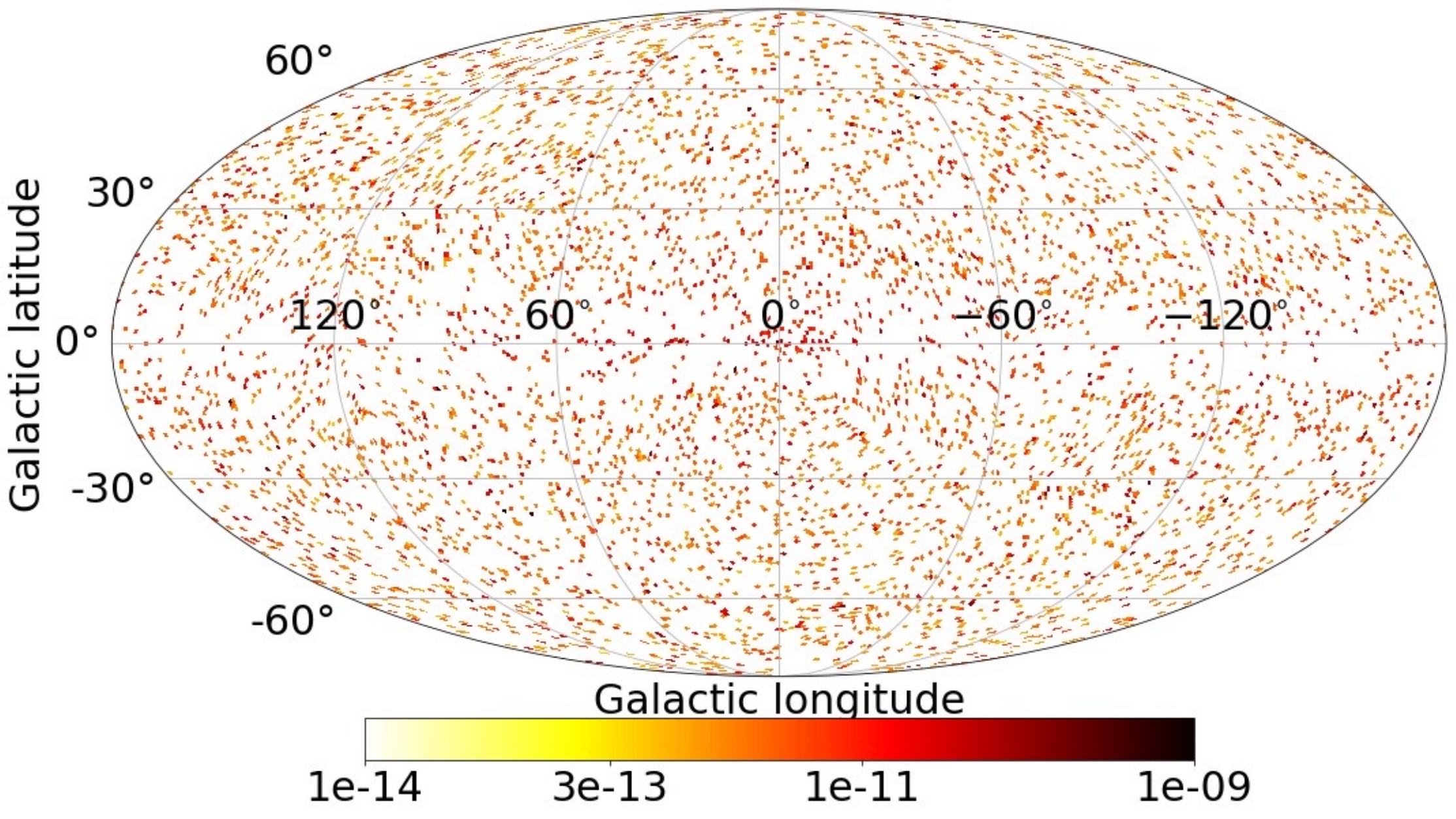}\,

    \caption{\label{4FGL_skymap}Skymap of the 4FGL-DR4 catalog after Galactic sources  and unassociated sources within $|b|<10^\circ$ are removed. The colorbar indicates the gamma-ray energy flux in erg cm$^{- 2}$ s$^{- 1}$. The dimmest source in the catalog has energy flux $3.09 \times 10^{-13}$ erg cm$^{- 2}$ s$^{- 1}$.}
\end{figure}

The 4FGL-DR4 catalog contains 7,195 sources, some of which are Galactic. Motivated by the fact that the UHECR sources are likely extragalactic \citep{Auger_dipole_2017}, we remove Galactic sources that are flagged as pulsars, supernova remnants, pulsar wind nebulae, globular clusters, binaries, star-forming regions, or novae. Many sources in the Fermi-LAT Galactic plane region are unassociated \citep{4FGL}, so in order to fully eliminate Galactic sources we must remove all unassociated low-latitude ($|b|< 10^\circ$) sources. However, to verify that this cut does not significantly affect our simulated dipole, we perform an additional test where we leave a randomly-selected set of low-latitude, unassociated sources chosen such that 1) the density per unit solid angle of low-latitude unassociated sources matches the density per solid angle of high-latitude unassociated sources and 2) the gamma-ray energy flux of all selected low-latitude unassociated sources falls within one standard deviation of the gamma-ray energy flux of the high-latitude unassociated sources. We confirm that there is no difference between this scenario and the one where all low-latitude unassociated sources are removed.
In total, we remove 1740 sources from the catalog, 1146 of which are unassociated. A skymap of the resulting source catalog is shown in Figure \ref{4FGL_skymap}.  

The 4FGL catalog has no redshift data provided. However, the catalog shares sources with the 4LAC-DR3 \citep{4LAC} and the 3FHL \citep{3FHL} catalogs, which have redshift data provided for many of their sources. We obtain the redshift information for 1805 sources from 4LAC-DR3 and 24 additional (not in 4LAC-DR3) sources from 3FHL. It is noteworthy that this provides redshift data for some of the nearest and brightest 4LAC and 3FHL sources, which are significant contributors to the large-scale anisotropy that we evaluate for simulated UHECRs.

\begin{figure}[t]
    \centering
    \includegraphics[width = 1\linewidth]{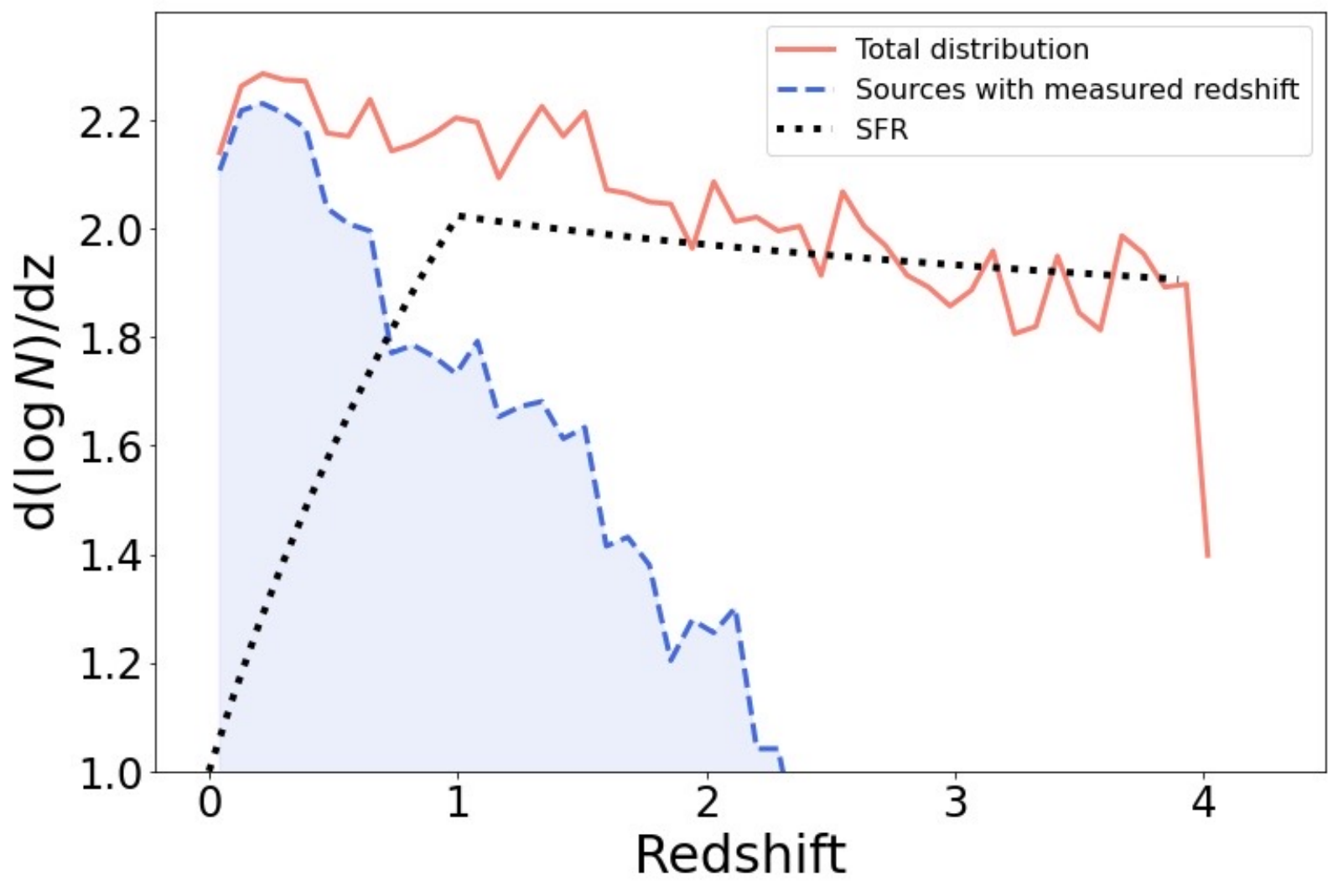}

    \caption{\label{redshifts} Redshift distribution of sources in 4FGL-DR4 after removing Galactic sources and low-latitude unassociated sources. Sources with known redshifts are indicated by the blue shaded region. Sources without distance information are assigned random redshifts following an SFR distribution \citep{SFR} as indicated by the black dotted curve. The sum of the two are shown as the red solid curve.  
    }
\end{figure}

For those sources that do not have redshift data provided in 4LAC-DR3 or 3FHL, we assign random redshifts following the star formation rate (SFR) as described in \citet{SFR}. Random redshifts are assigned up to a maximum redshift of $z = 4$; this maximum value is chosen because it is the maximum redshift of sources in the {\it Fermi}-LAT 3FHL and 4LAC-DR3 catalogs. The source distribution before and after random redshift assignment is shown in Figure \ref{redshifts}. 

We note that the sources with redshift data provided are typically those that are the nearest and brightest, while sources without measured redshift tend to comprise a more dim, distant background. We quantify the uncertainty in our results due to fluctuations in random redshift assignment by performing simulations for several different realizations of the random background, and we verify that random redshift assignment does not significantly change our results. Finally, as a comparison to evaluate the impact of our choice to assign redshift following an SFR distribution, we test a scenario where sources without redshift data provided are assigned redshift following a uniform distribution, and we conclude that this does not meaningfully affect our simulated dipole.

\section{Simulation\label{simulation}}

Without loss of generality, we consider two source weighting schemes. In the first scenario, we assume that the number of particles injected per source is weighted according to the gamma-ray energy flux at 100~MeV-100~GeV. We refer to this case as ``gamma-ray flux weighting". In the second case, we consider a scenario where the probability of detecting UHECRs at Earth coming from each source is the same. We call it ``uniform weighting”. This scenario is likely unphysical as it assumes that more distant sources are also more luminous. However, it can be justified because it accounts for the fact that at greater distances, the gamma-ray source catalogs are likely to be less complete. 
While neither of these weighting schemes are expected to be a perfect reflection of physical reality,  considering these two very different scenarios is intended to provide perspective on the UHECR dipole attainable with the catalogs of resolved gamma-ray sources. While we ran preliminary tests using some other commonly-considered weighting schemes, these weighting schemes yielded dipole amplitudes even larger than that found with the gamma-ray flux weighting, and were therefore redundant in the context of our results.

We simulate the propagation of cosmic rays from gamma-ray sources using  CRPropa~3.2 \citep{crpropa3.2, crpropa}. 
We use an injection spectrum following a power law and exponential cutoff at maximum rigidity:

$$
\frac{dN}{dE} = \begin{cases}
f_i E^{-\gamma}, & E \le ZR_{\text{max}} \\
f_i E^{-\gamma} e^{1 -E/ZR_{\text{max}}}, & E > ZR_\text{max}
\end{cases}
$$

We adopt a spectral index $\gamma = -1.82$ and maximum rigidity $R_{\text{max}} = 10^{18.15}$~V. We adopt luminosity fractions $I_{\mathrm{p}} = 0.620$, $I_{\mathrm{He}} = 0.233$, $I_{\mathrm{N}} = 0.533$, $I_{\mathrm{Si}} = 0.148$, and $I_{\mathrm{Fe}} = 0.024$ 
that fit the Auger unfolded spectrum reported in \citet{Auger_spectrum_2020} and the measurements of the depth of shower maximum reported in \citet{Auger_Xmax_2019} considering a scenario of sources distributed homogeneously and with the same luminosity. Using such a scenario results in a conservative estimate of the expected anisotropy, as a lighter composition or a higher rigidity cutoff would be expected to result in a stronger dipole \citep{di_Matteo_2018}. 

Our simulation accounts for the energy losses due to interactions with the cosmic microwave background and extragalactic background light \citep{IRB} via electron pair production, photodisintegration, and photopion production, as well as nuclear decay. 

We treat the extragalactic magnetic field (EGMF) as a random turbulent field with a Kolmogorov spectrum described by two parameters: root mean square value of the magnetic fields $B_{\text{RMS}}$, and maximum correlation length $l_{\text{coh}}$. In particular, we adopt $B_{\text{RMS}} \sqrt{l_{\text{coh}}} =  1\times10^{-13}$ G $\sqrt{\text{Mpc}}$. 
This choice is motivated by the upper bound on cosmological magnetic fields by \citet{jedamzik2019a} ($B_{\text{RMS}} \lesssim 10^{-11} \; \text{G}$). It is also in agreement with lower bounds derived from gamma-ray observations, which generally indicate $B_{\text{RMS}} \gtrsim 10^{-17} \; \text{G}$ at coherence lengths $l_{\text{coh}} \gtrsim 100 \; \text{kpc}$~\citep[e.g.,][]{fermi2018a, magic2023a}. 
Note that EGMF effects on UHECRs depend not only on $B_{\text{RMS}}$ but also on $l_{\text{coh}}$. However, limits on the coherence length are almost non-existent or very weak~\citep[e.g.,][]{alvesbatista2020a}.

The average deflection angle of a cosmic ray with rigidity $R$ propagating a distance $D$ in a magnetic field of rms strength $B$ and coherence length $\lambda$ is given by \citep{EGMF_deflection_angle}: 
\begin{equation}
\theta_{s} \approx 0.025^{\circ}\sqrt{\frac{D}{\lambda}}\bigg(\frac{\lambda}{10 Mpc}\bigg)\bigg(\frac{B}{10^{-11} G}\bigg)\bigg(\frac{R}{10^{20} V}\bigg)^{-1}. 
\end{equation}
The angular distribution of cosmic rays after averaging over time is given by a simple Gaussian: 

\begin{equation}\label{eqn:angularSeparation}
    P(\phi) = \frac{1}{\theta_{s}\sqrt{\pi}} e^{-\phi^2/\theta_{s}^2}. 
\end{equation}

We simulate cosmic ray propagation directly from source to observer, accounting for all interactions and energy losses that occur during propagation. We then 
assign the cosmic ray a deflection angle $\phi$ from the source following equation~\ref{eqn:angularSeparation} using its rigidity at injection. In addition to the deflection angle, the EGMF also introduces a time delay to particles traversing the field, $\tau \approx  {D\theta_{s}^2}/{(4c)}$, corresponding to a distance $D_{\text{add}} = {D\theta_{s}^2}/{4}$. To account for this additional travel distance, each particle is set back a distance $D_{\rm add}$ at its initial injection, allowing for any extra interactions and energy losses due to the time delay to occur.

To account for cosmic ray deflections in the Galactic magnetic field, we adopt the GMF model of \citet{JF12} and use the ``lensing" technique from \citet{crpropa}.  We include both the random striated and turbulent components of the field model. For the turbulent field, we use Kolmogorov spectral index 5/3, coherence length 60 pc, minimum and maximum scales of the turbulence 8 pc and 272 pc, respectively, and field strength following the parameters in Table 1 of \citet{JF12}. To generate the lens, we perform backtracking simulations of antiprotons with discrete rigidities on $\log_{10} R/\mathrm{V} \in [17.0, 21.0]$ and record their deflections in the GMF. These results are then combined to generate a Galactic lens. We apply the Galactic lens to simulated skymaps of UHECRs using CRPropa's \texttt{ParticleMapsContainer} feature, which yields a normalized skymap of UHECR events.  

\begin{figure}[t]
    \centering
    \includegraphics[width = 1\linewidth]{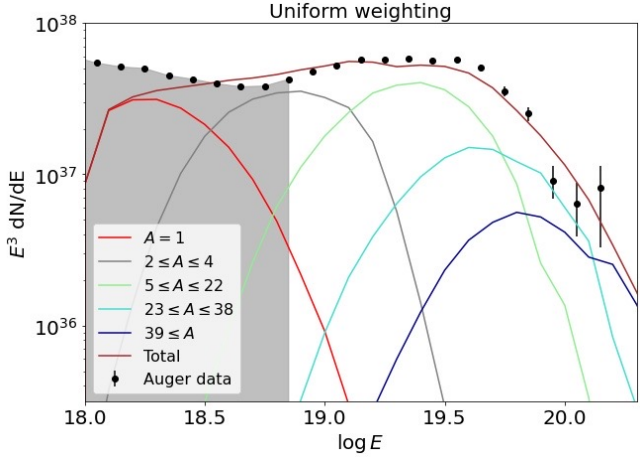}
    \caption{\label{spectra_uniform} Spectrum at Earth of simulated UHECRs from {\it Fermi}-LAT sources. The energy range below 8 EeV is shaded in gray as only events with energy greater than 8~EeV are used in evaluating the dipole as in  \citet{Auger_dipole_2017}.}
\end{figure}

The dipole amplitude and direction are evaluated using Rayleigh analysis following the method of the Pierre Auger Collaboration \citep{Auger_dipole_2017} and summarized in Appendix~\ref{appendix:rayleigh}. 

\section{Results \label{results}}
We present our simulation results using the 4FGL-DR4 catalog in Section~\ref{sec:resolved}. We then consider the effect of unresolved gamma-ray sources in Section~\ref{sec:unresolved}. Even though the uniform weight scenario is somewhat unphysical, we show the results from this case first. This is because, given the source distribution of the resolved gamma-ray sources, the uniform weighting scheme yields the best possible agreement with the dipole amplitude and composition measured by Auger. We show the results of the gamma-ray flux scenario, which has poorer agreement with the Auger observations, in Appendix \ref{appendix:eflux_weighting}.

\subsection{Simulated UHECRs from {\it Fermi}-LAT sources}\label{sec:resolved}

The simulated UHECR spectrum at Earth in the case of uniform source weighting is shown in Figure \ref{spectra_uniform}. The spectrum is in good agreement with the observed spectrum of Auger \citep{Auger_spectrum_2020}. The spectrum in the case of gamma-ray flux weighting is shown in Appendix \ref{appendix:eflux_weighting}. 

The dipole amplitudes and corresponding directions from the simulated UHECRs are summarized in Table \ref{results_table}. A flux map corresponding to the uniform weighting case is shown in Figure \ref{uniform_map}; a flux map in the case of gamma-ray flux weighting is shown in Appendix \ref{appendix:eflux_weighting}. The corresponding uncertainties are the standard deviations in the distribution of results from ten separate simulations with different iterations of random redshift assignment. 

\begin{figure}[t]
    \centering
    \includegraphics[width = 1\linewidth]{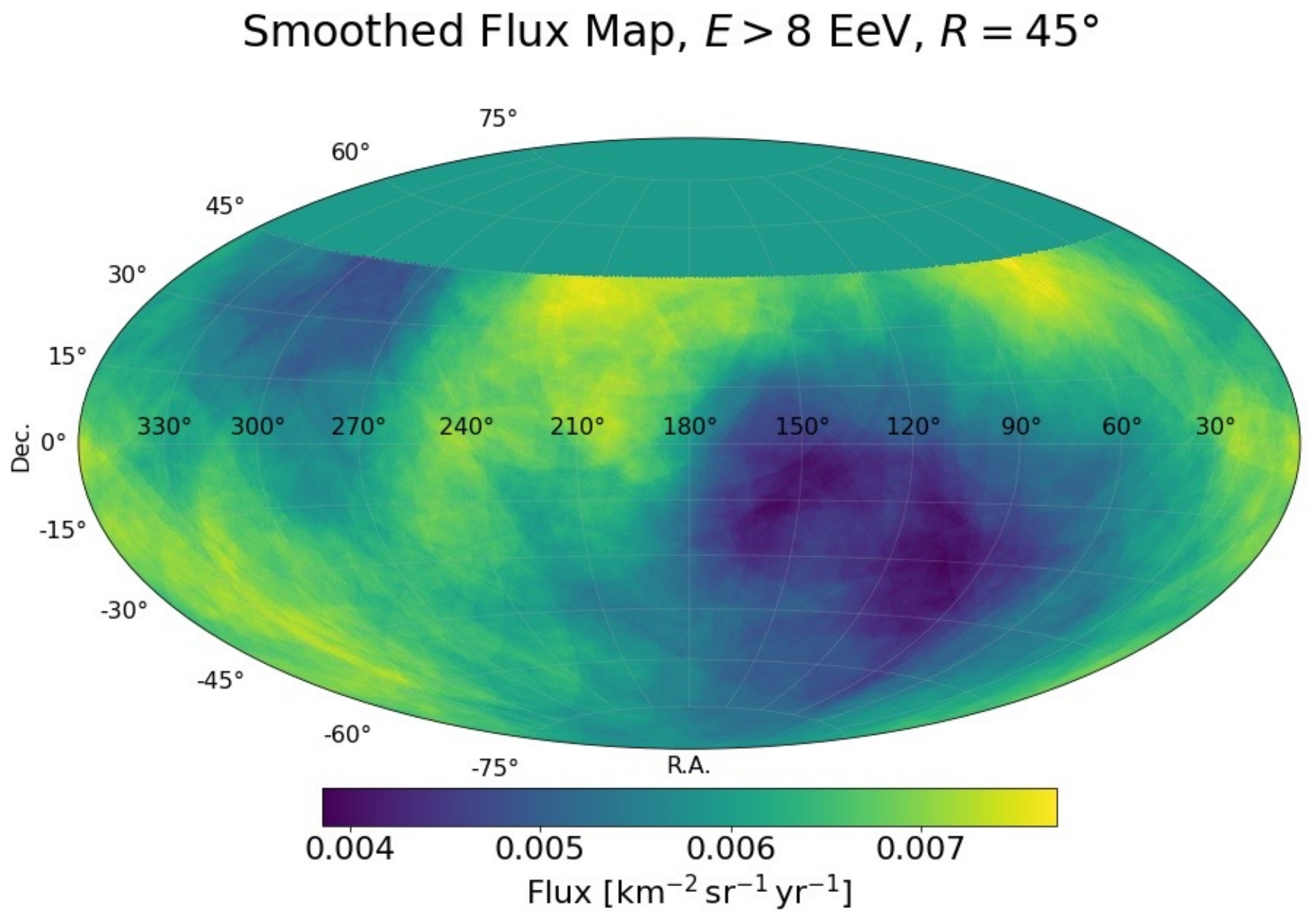}
    \caption{\label{uniform_map} Flux map corresponding to the case of uniform source weighting. The flux map is obtained by taking the ratio of the smoothed count density and exposure map following \citet{Auger_dipole_2017}. The skymap is smoothed with a $45^\circ$ top-hat function using the \texttt{healpy.sphtfnc.beam2bl} function.}
    
\end{figure}

\begin{table}[h]
    \begin{ruledtabular}
    \renewcommand{\arraystretch}{1.2}
    \begin{tabular}{cccc}
    Events $> 8$~EeV & Amplitude & $\alpha_d$ & $\delta_d$\\
    \hline
    Uniform & 33\% $\pm$7\% & $303^\circ \pm 17^\circ$ & $-11^\circ \pm 9^\circ$\\
    \hline
    Energy flux & 74\% $\pm$5\%  & $252^\circ \pm 7^\circ$  & $62^\circ \pm 3^\circ$\\
    \end{tabular}
    \end{ruledtabular}
    \caption{Dipole amplitude and direction for simulated UHECRs from 4FGL-DR4 sources in both weighting scenarios. The simulated datasets contain 32,000 events $>$ 8 EeV following \citet{Auger_dipole_2017}. 
    }
    \label{results_table}
\end{table}

In the case of gamma-ray flux weighting, we find a dipole amplitude of 74\%, which is more than ten times larger than that measured by Auger. In this case, the dipole is dominated by Markarian 421, an especially bright and relatively nearby source in the 4FGL-DR4 catalog. This disproportionately large effect by Markarian 421 is also noted in \citet{Bister_Mkn}. In the case of uniform source weighting, we evaluate a dipole amplitude of 33\%. This weighting scheme allows the more distant sources in the catalog to generate a stronger background against the nearby sources. However, even in this case, the dipole amplitude is significantly larger than the 6.5\% measured by Auger. In both scenarios, the dipole direction points to a different region of the sky from dipole direction measured by Auger. 

In both scenarios, the very large dipole amplitude indicates that the density of resolved gamma-ray sources is too low to generate a dipole as low as that measured by Auger. This disagreement is further confirmed by the tension between the simulated and observed dipole directions. Our results  imply that, given the magnetic field models used in this analysis, the resolved gamma-ray sources are insufficient to account for the population of UHECR sources producing the Auger dipole. In other words, there must be a population of UHECR sources that are unobserved by the current gamma-ray facilities, and are gamma-ray dim.

\subsection{Unresolved Sources}\label{sec:unresolved}

As a final test, we consider the effect that unresolved gamma-ray sources could have on the simulated cosmic ray dipole. We test unresolved source populations using two approaches. In the first approach, we infer an unresolved source population using a source count distribution and consider the UHECR dipole when these sources are weighted according to their energy flux. In the second approach, we generate a fixed number of unresolved sources that are weighted uniformly. In both cases, the unresolved sources are assigned positions following an isotropic distribution and random redshifts following the SFR distribution discussed previously. While it may be possible to distribute the unresolved source positions anisotropically such that the direction of the Auger dipole is reproduced, our assumption of an isotropic distribution is a simple, conservative approximation to yield the lowest possible dipole with the fewest number of additional sources. The unresolved sources are appended to the 4FGL-DR4 catalog, and we evaluate the dipole of cosmic rays simulated from this new distribution.

For the case in which unresolved sources are weighted by energy flux, we base our unresolved source population on the source count distribution from \citet{unresolved_sources}. This work uses machine learning to derive a source count distribution following $\frac{dN}{dS} \propto S^{-2}$ below the 4FGL-DR4 sensitivity of $2 \times 10^{-10}$ ph cm$^{-2}$ s$^{-1}$. Following this distribution, we generate a population of unresolved sources extending to the minimum luminosity of $S_{\rm min} = 5 \times 10^{-12}$ ph cm$^{-2}$ s$^{-1}$ used in the analysis of \citet{unresolved_sources}. We derive the energy flux of each source from its photon flux assuming a source energy spectrum $\frac{dN_{\gamma}}{dE_\gamma} \propto E_\gamma^{-b_\gamma}$ with $b_\gamma = 2.23$, which is the median spectral index of 4FGL-DR4 extragalactic sources. We find that in this case, there is no measurable difference compared to the energy flux weighted simulation without the additional population of unresolved sources. This is due to the fact that, despite the very large number of added sources, the very low energy fluxes of the unresolved sources are not sufficient to balance out the effect of a bright source like Markarian 421.

In the second test, we incrementally add fixed numbers of unresolved sources to the 4FGL-DR4 catalog, and we weight all sources uniformly. In this case, we find that a number of $\sim80,000$ added sources (corresponding to a local source density of $\sim 5\times 10^{-4}$ Mpc$^{-3}$) is required to lower the dipole amplitude to a level analogous to that measured by Auger. This result is informative because it demonstrates that to have a dipole amplitude as low as that measured by Auger, the UHECR source density must be significantly larger than the resolved gamma-ray source population, especially the population of sources comprising the UHECR background. The result is also in agreement with other work (e.g., \citealp{bister2023constraints}) that has derived a similar source density necessary to achieve a dipole as low as Auger's.

The result of this test with added unresolved sources strengthens our conclusion that UHECR injection cannot be proportional to gamma-ray energy flux, and in fact that gamma-ray-dim sources must account for a substantial portion of UHECR accelerators.   
Assuming that the accepted EGMF and GMF models do not severely underestimate average magnetic field strengths, a dipole amplitude as low as Auger's can be achieved only with a very high source density, which the resolved gamma-ray sources are not sufficient to account for.

\section{Summary and Discussion \label{summary}}

In this work, we simulate UHECRs originating from extragalactic {\it Fermi}-LAT gamma-ray sources and compare the simulated dipole to that measured by Auger. We use the {\it Fermi}-LAT 4FGL-DR4 catalog as the basis for our simulations.
We find that the resolved gamma-ray sources are insufficient to account for the population of sources producing UHECRs $>$ 8 EeV, primarily because of low density of nearby sources. We conclude that there must exist a population of UHECR sources that are gamma-ray dim. 

Our assumption of a turbulent extragalactic magnetic field (EGMF) pervading the universe is not realistic. Ideally, the distribution of magnetic fields in the large-scale structure of the universe should be taken into account. In this case, cosmic voids would be the dominant component driving UHECR deflections~\citep{alvesbatista2017c}. By adopting a somewhat strong EGMF ($B_{\text{RMS}} \sqrt{l_{\text{coh}}} = 10^{-13} \; \text{G} \sqrt{\text{Mpc}}$), we are conservatively choosing a scenario that increases overall deflections and thus reduces the dipole amplitude, which strengthens our conclusions, in qualitative agreement with \citet{dundovic2019a} and \citet{hackstein2018a}.

As an additional test of the effect of EGMF strength on our simulated dipole, we also consider a larger EGMF strength of $B_{\text{RMS}} \sqrt{l_{\text{coh}}} = 0.16  \text{nG}$. While this value exceeds the upper bound on cosmological magnetic fields by \citet{jedamzik2019a}, it was obtained as the best-fit value in \citet{vanVliet_2021}, which uses simulations of UHECRs from nearby star-forming galaxies to reproduce the intermediate-scale anisotropy observed by Auger. In both energy flux and uniform-weight scenarios, the larger magnetic field strength does not alter the dipole amplitude sufficiently to change the conclusion of our work. In the uniform weight case, the order of magnitude change in $B_{\text{RMS}}$ only lowers the dipole by a few percent. In the energy flux weight case, the larger EGMF strength actually raises the dipole by $\sim$ 20\%, this effect caused by the effect of EGMF on the balance between the brightest source, Markarian 421, and the other background sources.

 While the JF12 model of the GMF has been found to fit many observable parameters well, the model is still not perfectly constrained, and it is expected that new GMF models will continue to improve our understanding of UHECR deflection and source searches in the future \citep{boulanger2018a, coleman2023a}. Other works \citep{vanVliet_2021, bister2023constraints} using the JF12 model make the assumption that variations in the GMF model can affect the amplitude and direction of the measured dipole somewhat, but that the overall trend of the dipole's behavior given a particular source distribution and EGMF strength would not change drastically. In the case of this work, while future models of the GMF could lower the evaluated dipole amplitude somewhat, a new GMF model would have to employ significantly larger magnetic field strengths to bring our evaluated dipole amplitude into agreement with that measured by Auger. The recent work \citet{new_JF12} provides updated models of the GMF, but no model in the new ensemble differs enough from JF12 to significantly lower the expected dipole amplitude.

In general, our finding is in agreement with previous works which have shown that it is necessary to have a very high density of nearby sources to produce a dipole amplitude as low as that observed by Auger \citep{vanVliet_2021, 2021ApJ...913L..13D, bister2023constraints, Bister_Mkn}. 
Given our assumptions about the extragalactic and Galactic magnetic fields, our result shows that the density of UHECR sources must be larger than that of the resolved gamma-ray sources. 

Future neutrino telescopes \citep{snowmass_white} will seek to observe ultra-high-energy (UHE) neutrinos and to identify their sources. UHE neutrinos can be produced in UHECR sources (e.g., \citealp{Fang:2013vla}), and their production is accompanied by that of UHE gamma rays. The UHE gamma rays may cascade down to GeV energies by the radiation background of the source or the extragalactic background lights. Our result implies that most UHECR accelerators might lack the conditions necessary for significant hadronic interactions to occur. Such sources may not only be gamma-ray dim, but also dim emitters of UHE neutrinos.

\begin{acknowledgements}
The work of A.P. and K.F. is supported by the Office of the Vice Chancellor for Research and Graduate Education at the University of Wisconsin-Madison with funding from the Wisconsin Alumni Research Foundation. K.F. acknowledges support from National Science Foundation (PHY-2110821, PHY-2238916) and NASA (NMH211ZDA001N-{\it Fermi}). This work was supported by a grant from the Simons Foundation (00001470, KF). 
R.A.B. is funded by: the ``la Caixa'' Foundation (ID 100010434) and the European Union's Horizon~2020 research and innovation program under the Marie Skłodowska-Curie grant agreement No~847648, fellowship code LCF/BQ/PI21/11830030; grants PID2021-125331NB-I00 and CEX2020-001007-S, funded by MCIN/AEI/10.13039/501100011033,
by ``ERDF A way of making Europe'', and the MULTIDARK Project RED2022-134411-T.
\end{acknowledgements}

\software{CRPropa3.2 \citep{crpropa3.2}. We have made the code for converting CRPropa output to events observed by Auger publicly available at \url{https://github.com/angelinapartenheimer/CRDipy.git}}

\appendix

\section{Spectrum at Earth and flux map in the case of gamma-ray flux weighting} \label{appendix:eflux_weighting}

Here we present the flux map and spectrum at Earth of simulated cosmic rays in the case of gamma-ray flux weighting, which yields a simulated dipole amplitude of 74\%. In this case, the UHECR dipole is dominated by Markarian 421, an especially bright and nearby source in the 4FGL-DR4 catalog. The effect of Markarian 421 is demonstrated by the bright spot in the center of the flux map. Because the UHECR signal is dominated by one source, the simulated UHECR spectrum deviates slightly from the Auger data at high energies. However, this small amount of disagreement cannot explain the order of magnitude in difference between the simulated dipole and the 6\% dipole observed by Auger. 

\begin{figure}[h]
    \centering
    \includegraphics[width = 1\linewidth]{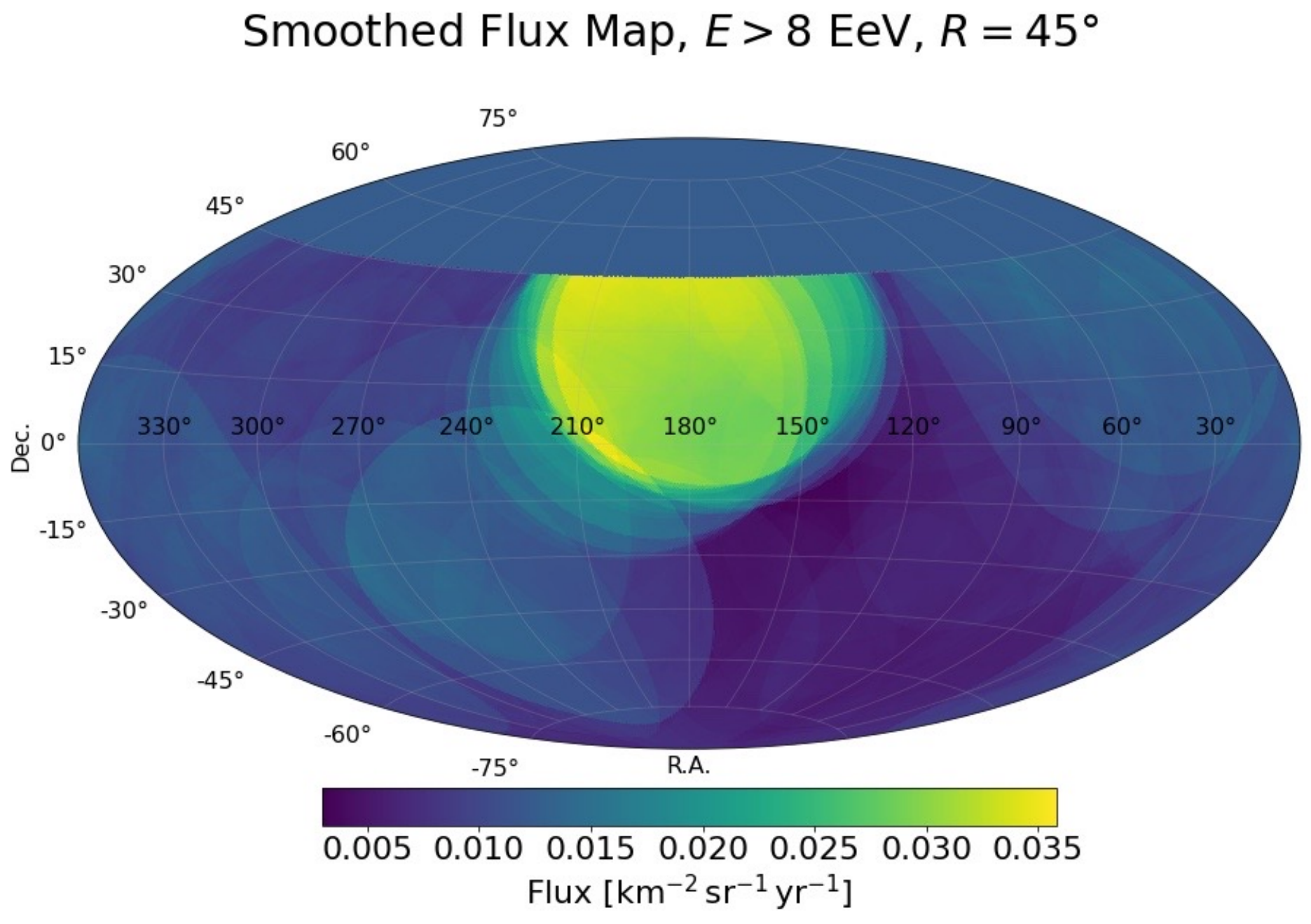}
    \caption{\label{flux_map} Flux map corresponding to the case of gamma-ray flux weighting. The flux map is obtained by taking the ratio of the smoothed count density and exposure map following \citet{Auger_dipole_2017}. The skymap is smoothed with a $45^\circ$ top-hat function using the \texttt{healpy.sphtfnc.beam2bl} function.}  
\end{figure}

\begin{figure}[t]
    \centering
    \includegraphics[width = 1\linewidth]{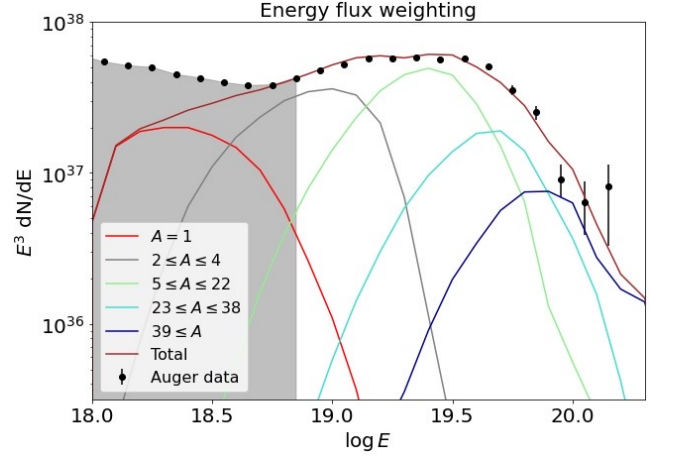}
    \caption{\label{spectrum_eflux} Spectrum at Earth of simulated UHECRs from {\it Fermi}-LAT sources in the case of gamma-ray flux weighting. Because the signal is dominated by one source, there is some discrepancy between the simulated spectrum and Auger data at high energies; however, the disagreement in the spectra is not sufficient to explain the order of magnitude in difference between the simulated and observed dipole amplitude.}
\end{figure}

\section{Rayleigh analysis}\label{appendix:rayleigh}

This section recapitulates the Rayleigh analysis method outlined in \citet{Auger_dipole_2017}. 

The dipole is calculated in terms of its components perpendicular $d_\perp$ and parallel $d_z$ to the Earth's rotation. The complete dipole is given by
\begin{equation}
d = \sqrt{d_{\perp}^2+d_{z}^2}. 
\end{equation}

To reconstruct the component of the dipole perpendicular to Earth's rotation $d_\perp$, we perform Rayleigh analysis in right ascension $\alpha$. Applying the Galactic magnetic lens returns a density map of cosmic ray events. Each position in the sky (right ascension $\alpha_i$, declination $\delta_i$) has a corresponding weight $w_i$ corresponding to the event density at that position. The first-harmonic Fourier components are:

\begin{equation}
    a_{\alpha} = \frac{2}{\mathcal{N}}\displaystyle\sum_{i = 1}^{N_{pix}} w_{i}\cos{\alpha_i},\hspace{0.5cm} b_{\alpha} = \frac{2}{\mathcal{N}}\displaystyle\sum_{i = 1}^{N_{pix}} w_{i}\sin{\alpha_i}    
\end{equation}

where $\mathcal{N} = {\displaystyle\sum_{i = 1}^{N_{pix}}w_i}$ and $N_{pix}$ is the number of pixels in the skymap. 

The amplitude and phase of the first harmonic of modulation are: 
\begin{equation}
r_\alpha = \sqrt{a_{\alpha}^2 + b_{\alpha}^2}, \hspace{0.5cm}
\tan{\phi_\alpha} = \frac{b_{\alpha}}{a_\alpha}. 
\end{equation}
The dipole component $d_\perp$ is then given by: 
\begin{equation}
d_\perp \approx \frac{r_\alpha}{\langle \cos{\delta} \rangle}
\end{equation}
where 
$\langle \cos{\delta} \rangle = \frac{1}{\mathcal{N}}\displaystyle\sum_{i = 1}^{N_{pix}} \cos{\delta_i}$ is the mean cosine of declinations of detected events. The corresponding right ascension of the dipole is: 
\begin{equation}
\alpha_d = \phi_\alpha. 
\end{equation}

To reconstruct the dipole component parallel to the Earth's rotation $d_z$, we perform Rayleigh analysis in azimuthal angle $\phi$. This yields first-harmonic Fourier components $a_\phi$ and $b_\phi$. In this case, the dipole component $d_z$ is given by: 
\begin{equation}
d_z \approx \frac{b_\phi}{\cos{l_{obs}}\langle\sin{\theta}\rangle}
\end{equation}
where $l_{obs}$ is the latitude of the Pierre Auger observatory, and $\langle\sin{\theta}\rangle$ is the mean sign of event zenith angles. The declination of the dipole is: 
\begin{equation}
\tan{\delta_d} = \frac{d_z}{d_\perp}.
\end{equation}

\section{Results from 3FHL and 4LAC-DR3 Sources}\label{appendix:othergammaCat}

We test the dipole amplitude and direction of UHECRs coming from 3FHL or 4LAC-DR3 sources alone. The 3FHL catalog provides sources observed by {\it Fermi}-LAT in the range 10 GeV--2 TeV from 7 years of data, and is the first {\it Fermi}-LAT catalog to use Pass 8 event-level analysis \citep{3FHL}. It contains 1556 sources. The 4LAC catalog gives data on the high-latitude AGNs on 50 MeV - 1 TeV \citep{4LAC}, and it is a subset of the 4FGL-DR3 catalog. It contains 2863 sources. The 4LAC-DR3 catalog contains no Galactic sources; the 3FHL catalog contains some Galactic sources, and these are removed following the same method outlined in Section \ref{gr_sources}. This results in 197 sources being removed from 3FHL, 72 of which are unassociated. 3FHL has redshift data for 35\% of its extragalactic sources; 4LAC-DR3 contains redshift data for 63\% of its sources. The sources without redshift data provided are assigned random redshifts following the SFR distribution from \citet{SFR}.  

We perform separate simulations for UHECRs from the 3FHL and 4LAC-DR3 sources. In each case, both energy flux and uniform source injection weighting are tested. Results from the tests are shown in Table \ref{3fhl_4lac_results}. In all cases, the dipole amplitude is significantly larger that observed by Auger. 

\begin{table}
    \begin{ruledtabular}
    \renewcommand{\arraystretch}{1.2}
    \begin{tabular}{cccc}
    8 EeV $<$ & Dipole amplitude & $\alpha_d$ & $\delta_d$\\
   \hline
    3FHL flux weight & 128\% & $205^\circ$ & $61^\circ$\\
    \hline
    3FHL uniform weight & 71\% & $295^\circ$ & $39^\circ$\\
    \hline
    4LAC flux weight & 77\% & $253^\circ$ & $70^\circ$\\
    \hline
    4LAC uniform weight & 31\% & $303^\circ$ & $-28^\circ$\\
    \end{tabular}
    \end{ruledtabular}
    \caption{Dipole of UHECRs from separate simulations of 3FHL and 4LAC-DR3 sources.}
\label{3fhl_4lac_results}
\end{table}

\end{document}